\documentclass[aps,prl,preprintnumbers,twocolumn,floatfix,superscriptaddress,amsmath,amssymb,amsfonts]{revtex4-2}
\usepackage{graphicx}
\usepackage{bm}
\usepackage{textcomp}
\usepackage{multirow}
\usepackage[colorlinks=true,linkcolor=blue,citecolor=blue,urlcolor=blue]{hyperref}
\usepackage{tabularx}
\usepackage{gensymb}
\usepackage{color}
\usepackage{array}
\usepackage{hhline}
\usepackage[dvipsnames]{xcolor}
\usepackage[normalem]{ulem}
\usepackage{verbatim}   % useful for program listings
\usepackage{subfigure}  % use for side-by-side figures
\usepackage{braket}
\usepackage{float}
\DeclareGraphicsExtensions{.ps,.eps,.pdf}
\DeclareGraphicsExtensions{.jpg,.png}

\begin{document}

\title{From hierarchical triangular spin liquid to multi-\texorpdfstring{$q$}{q} spin texture in spinel GeFe\texorpdfstring{$_2$}{2}O\texorpdfstring{$_4$}{4}}

\author{L. Chaix} \email[\textit{Corresponding author:} ]{laura.chaix@neel.cnrs.fr}
\affiliation{Institut Néel, CNRS and Université Grenoble Alpes, 38042 Grenoble, France}

\author{J. Robert}
\affiliation{Institut Néel, CNRS and Université Grenoble Alpes, 38042 Grenoble, France}

\author{E. Chan}
\affiliation{Institut Néel, CNRS and Université Grenoble Alpes, 38042 Grenoble, France}

\author{E. Ressouche} 
\affiliation{Université Grenoble Alpes, CEA, IRIG, MEM, MDN, 38000 Grenoble, France}

\author{S. Petit}
\affiliation{Université Paris-Saclay, CNRS, CEA, Laboratoire Léon Brillouin, 91191 Gif-sur-Yvette, France}

\author{C. V. Colin}
\affiliation{Institut Néel, CNRS and Université Grenoble Alpes, 38042 Grenoble, France}

\author{R. Ballou}
\affiliation{Institut Néel, CNRS and Université Grenoble Alpes, 38042 Grenoble, France}

\author{J. Ollivier}
\affiliation{Institut Laue-Langevin, 71 Avenue des Martyrs, 38042 Grenoble, France}

\author{L.-P. Regnault} 
\affiliation{Université Grenoble Alpes, CEA, IRIG, MEM, MDN, 38000 Grenoble, France}

\author{E. Lhotel}
\affiliation{Institut Néel, CNRS and Université Grenoble Alpes, 38042 Grenoble, France}

\author{V. Cathelin}
\affiliation{Institut Néel, CNRS and Université Grenoble Alpes, 38042 Grenoble, France}

\author{S. Lenne}\altaffiliation{\textit{Current address:} School of Physics, CRANN, Trinity College, Dublin 2, Ireland}
\affiliation{Institut Néel, CNRS and Université Grenoble Alpes, 38042 Grenoble, France}

\author{C. Cavanel}
\affiliation{Institut Néel, CNRS and Université Grenoble Alpes, 38042 Grenoble, France}

\author{F. Damay}
\affiliation{Université Paris-Saclay, CNRS, CEA, Laboratoire Léon Brillouin, 91191 Gif-sur-Yvette, France}

\author{E. Suard}
\affiliation{Institut Laue-Langevin, 71 Avenue des Martyrs, 38042 Grenoble, France}

\author{P. Strobel}
\affiliation{Institut Néel, CNRS and Université Grenoble Alpes, 38042 Grenoble, France}

\author{C. Darie}
\affiliation{Institut Néel, CNRS and Université Grenoble Alpes, 38042 Grenoble, France}

\author{S. deBrion}
\affiliation{Institut Néel, CNRS and Université Grenoble Alpes, 38042 Grenoble, France}

\author{V. Simonet}
\affiliation{Institut Néel, CNRS and Université Grenoble Alpes, 38042 Grenoble, France}

\date{\today}

\begin{abstract}
Combining macroscopic measurements, neutron scattering and modeling, we identify in the GeFe$_2$O$_4$ spinel a correlated paramagnetic state resulting from the predominance of third-neighbor antiferromagnetic interactions. These interactions materialize 4 isolated families of triangular planes with 120$^{\circ}$ spins emerging from the underlying pyrochlore lattice. At lower temperatures, a phase transition occurs from this hierarchical spin liquid to a non-coplanar spin texture that is characterized by 6 propagating vectors. This unusual multi-$q$ order is triggered by the presence of weaker interactions up to the sixth neighbors. The system is remarkably successful in coupling the different triangular planes while maintaining their two-dimensional 120$^{\circ}$ order. Our study highlights the hierarchy of interactions involved in GeFe$_2$O$_4$, which is singular among spinel compounds since first-neighbor interactions are only a small fraction of the dominant third neighbor ones. 
\end{abstract}

%\pacs{PACS }
%\keywords{}

\maketitle

Magnetic frustration in condensed matter reveals new properties and states of matter beyond conventional magnetic behaviors \cite{Moessner2006,Normand2009,Balents2010}. The archetypal corner-sharing tetrahedron-based pyrochlore lattice for instance can promote remarkable ground states devoid of long-range order due to geometrically frustrated interactions. These include classical or quantum spin liquids \cite{Reimers1992,Canals1998,Moessner1998,Lee2002,Gardner1999} for antiferromagnetic (AFM) interactions and spin ices for effective ferromagnetic (FM) interactions \cite{Gingras_McClarty_2014,Udagawa2021}. They have been predicted/observed in pyrochlore oxides with magnetic rare-earths where first neighbor interactions are a major ingredient \cite{Gardner2010}. A rich physics also arises when interactions from further neighbor compete with first neighbors ones. This is made possible in spinel compounds, of formula $AM_2$O$_4$, where the pyrochlore lattice is occupied by transition metal elements $M$ with stronger exchange interactions and additional super-exchange paths \cite{Tsurkan2021,Lee2010}. Interesting behaviors were reported in Cr spinels: below a correlated paramagnetic region featuring a spin liquid \cite{Lee2002}, complex magnetic structures are observed facilitated by structural distortion that releases frustration and stabilizes magnetization plateaus under magnetic field \cite{Ueda2006}. Their long-range orders \cite{Chung2005,Chern2006,Gao2018}, often including several propagation vectors, are challenging to resolve, reflecting the complexity of the Hamiltonians. Interestingly, the non-magnetic $A$ element plays a role in the nature of first-neighbor interactions (combining direct and super-exchange interactions of opposite sign) then in the diversity of the behaviors \cite{Yaresko2008}. An appealing issue concerns the ultimate possibility of cancelling the first neighbor interactions in order to unveil novel physics from further neighbor ones. 

\begin{figure}
\includegraphics[width=1\columnwidth]{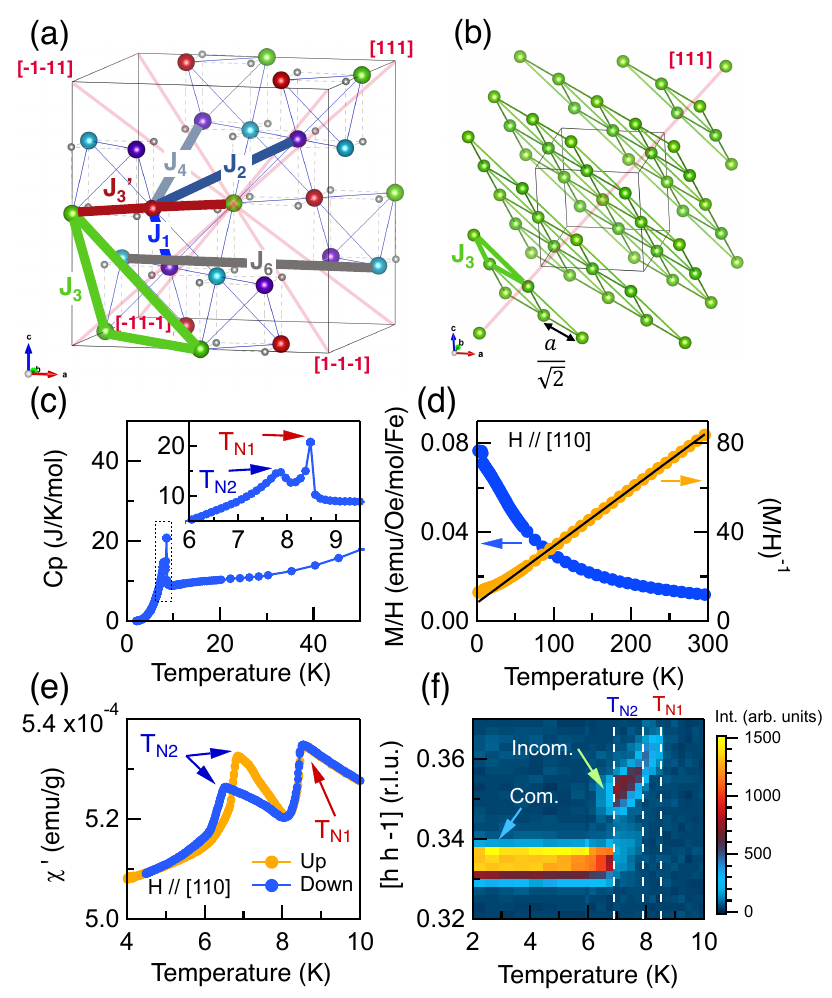}
\caption{(a) Fe$^{2+}$ pyrochlore lattice including oxygens (grey circles) in GFO spinel. The four Fe Bravais sublattices are shown with different colors and the magnetic interactions with colored lines. The four diagonals of the cube are indicated with red lines. (b) One Fe sublattice forming triangular planes perpendicular to the $[111]$ direction for which the relevant interaction is $J_3$. (c) Specific heat $vs$ temperature with the low temperature zoom in the inset. (d) $M/H$ $vs$ temperature measured in a magnetic field of $1$ T applied along $[110]$ (blue points). Its inverse (orange points) is fitted by a Curie-Weiss law $H/M=(T-\theta_{\text{CW}})/C$ (black line), with the Curie constant $C$ and the Curie-Weiss temperature $\theta_{\text{CW}}$. (e) Real part of the AC susceptibility measurements at 5.7 Hz through cooling and heating with an AC field of $0.55$ mT applied along $[110]$. (f) Single crystal neutron diffraction intensity map gathering $Q$-scans along the [$h$ $h$ $\bar{1}$] direction at different temperatures. Commensurate and incommensurate peaks are indicated with blue and green arrows. T$_{N1}$ and T$_{N2}$ are indicated by red and blue arrows or dashed lines in the different panels.}
\label{FIGURE1}
\end{figure}

We investigate this question in the less studied $A$Fe$_2$O$_4$ family \cite{Kamazawa1999,Kamazawa2003,Kamazawa2004,Dronova2024}, where first neighbor interactions are indeed inferred to change sign when moving from the Cd (AFM) to the Zn (FM) compound. We particularly turn to the germanate member, GeFe$_2$O$_4$ (GFO), noting that unlike in the $A$ = Cd, Mg or Zn iron spinels where Fe is trivalent, in the $A$ = Ge iron spinel GeFe$_2$O$_4$, Fe is divalent, reducing the direct exchange interactions \cite{Goodenough}. A transition towards a magnetic order through a double magnetic transition has been reported in powder GFO from macroscopic measurements and neutron diffraction experiments \cite{Perversi2018,Barton2014}. Perversi {\it et al.} proposed an amplitude modulated magnetic arrangement arising from strong spin/orbital frustration produced by first neighbour AFM interactions \cite{Perversi2018}. However, this structure is hardly compatible with earlier M\"ossbauer measurements showing that an identical molecular field acts on all Fe sites \cite{Imbert1966,Hartmann1968} and with the fact that third neighbour interactions should be dominant \cite{Plumier1966,Sandemann2023}.

In this article, we therefore revisit the magnetic order of GFO, combining modeling and experiments: macroscopic measurements, neutron diffraction (ND) and inelastic neutron scattering (INS) on a single crystal. We show that the magnetic order is very exotic: non-coplanar and described by 6 propagation vectors. This unconventional ordering is enabled by the strong hierarchy of interactions. Remarkably, it arises from a spin liquid stabilized at intermediate temperature by the leading third neighbor interactions, which build isolated triangular planes out of the pyrochlore lattice.

In the cubic cell of the spinel structure, the Fe$^{2+}$ magnetic ions are decomposed in 4 Bravais lattices at the corners of connected tetrahedra forming the pyrochlore lattice [red, green, blue and purple Fe atoms in Fig. \ref{FIGURE1}(a)]. Different magnetic exchange paths between Fe$^{2+}$ up to the sixth neighbors ($J_1$, $J_2$, $J_3$/$J_3'$, $J_4$ and $J_6$) are considered [Fig. \ref{FIGURE1}(a)]. There are actually two third-neighbor interactions corresponding to Fe-Fe distance of $\frac{a}{\sqrt{2}}\approx5.95 \text{ \AA}$, one going through a Fe atom ($J_3'$) and one without intermediate atom ($J_3$) \cite{Matsuda2008,Tomiyasu2011}. Considering solely the $J_3$ connectivity, each Fe sublattice forms triangular planes isolated from each other and stacked along the local $\langle 111 \rangle$ directions (cube's diagonals) [Fig. \ref{FIGURE1}(b)].

\begin{figure*}
\resizebox{17cm}{!}{\includegraphics[width=1.0\columnwidth]{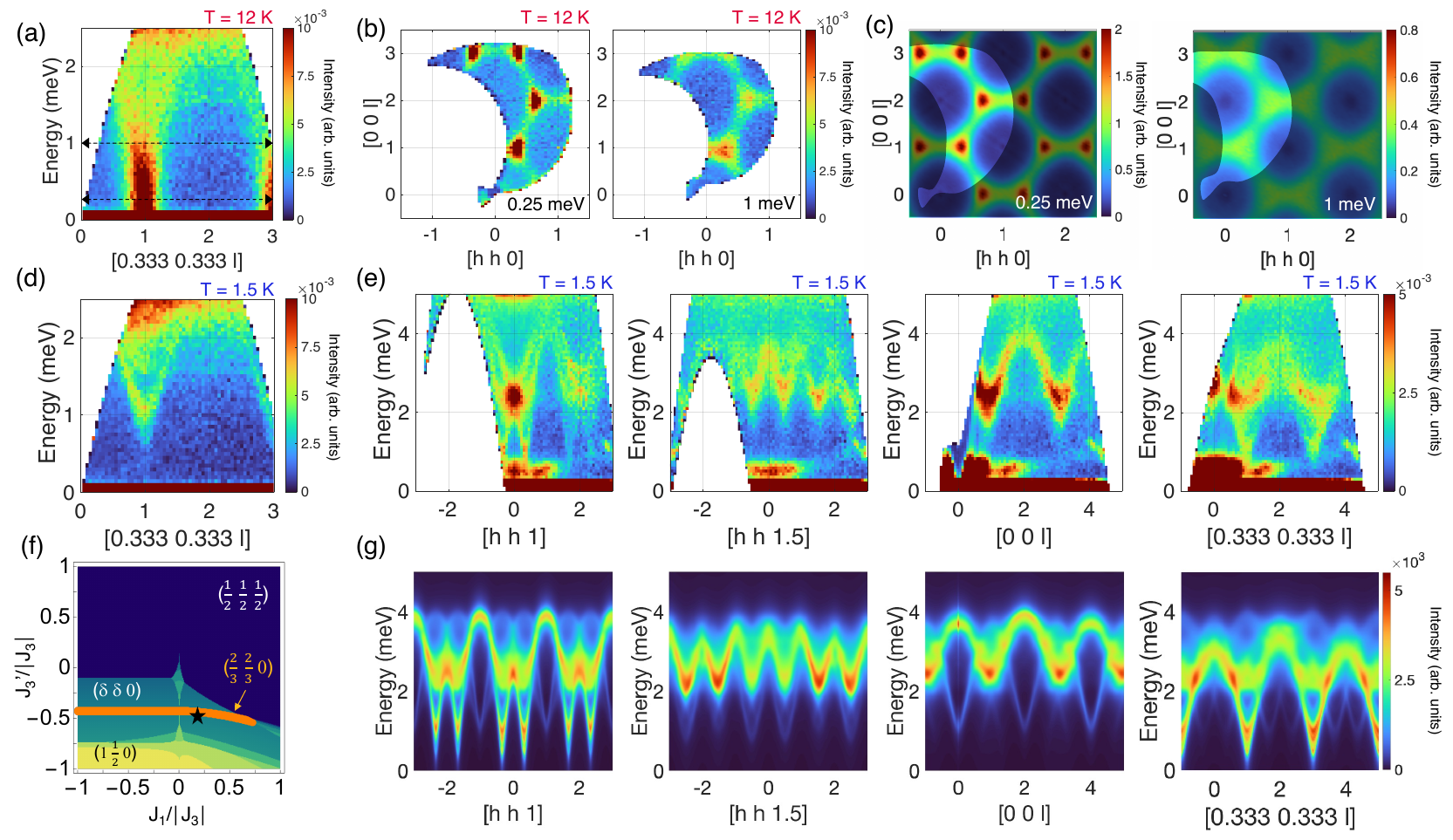}}
\caption{(a) INS energy-momentum map recorded with $\lambda$ = 4.8 $\text{\AA}$ at $T$ = 12~K above T$_{N1}$. (b) Corresponding constant energy-cuts in the [00$l$]-[$hh$0] scattering plane at 12~K and different energies indicated by horizontal black arrows in panel (a). (c) Monte-Carlo calculations of the dynamical structure factor in the correlated paramagnetic state using the $J_3$-model described in the text plotted as the experimental data of panel (b). (d) INS energy-momentum map recorded with $\lambda$ = 4.8 $\text{\AA}$ at $T$ = 1.5~K showing the gap in the ordered state. (e) INS energy-momentum maps for different $Q$ directions recorded with $\lambda$ = 3.4 $\text{\AA}$ at $T$ = 1.5~K. (f) $J_1$-$J_3'$ phase diagram with fixed dominant $J_3$ (AFM), $J_6$/$\lvert J_3 \lvert$ = -0.2 and $J_2$ = $J_4$ = 0 as well as an easy-plane anisotropy of $g_{\parallel}$/$g_{\perp}$ = 0.5. Colors represent phases with different propagation vectors. The ($\frac{2}{3}$ $\frac{2}{3}$ 0) region is shown in orange and the black star indicates the set of parameters used for the spin wave simulations of panel (g). (g) Spin wave dispersion for different $Q$ directions calculated with a dominant AFM $J_3$, $J_3'$/$\lvert J_3 \lvert$ = -0.5, $J_6$/$\lvert J_3 \lvert$ = -0.2, $J_1$/$\lvert J_3 \lvert$ = 0.2, $J_2$ = $J_4$ = 0, an easy-plane anisotropy $g_{\parallel}$/$g_{\perp}$ = 0.5 and all $Q$-domains averaged. Negative (positive) interactions correspond to AFM (FM).}
\label{FIGURE2}
\end{figure*}

We performed our measurements on a Ge$_{1-x}$Fe$_{2+x}$O$_4$ single crystal of 35.5(1) mg presenting a very weak inversion disorder $x$ = 0.057 (see Supplemental Material \cite{SupplementalMaterial}). At high temperatures, the magnetization measured with $H$~$\parallel$~$[110]$ shows a Curie-Weiss behavior with $\theta_{\text{CW}} = -31.9(3)$~K [Fig. \ref{FIGURE1}(d)], evidencing dominant AFM interactions and an effective magnetic moment $\mu_{\rm eff}=5.58(1)$ $\mu_B$. This value is in agreement with our crystalline electric field (CEF) analysis \cite{SupplementalMaterial}. Below 40~K, deviations to the Curie-Weiss law are observed in the magnetization [Fig. \ref{FIGURE1}(d)] and a broad bump around 20~K is detected in the specific heat [Fig. \ref{FIGURE1}(c)], indicative of the onset of short range magnetic correlations.  

The system orders through two magnetic transitions observed at T$_{N1}$ = 8.5~K and T$_{N2}$ = 7.9~K from sharp anomalies in the specific heat data [Fig. \ref{FIGURE1}(c)]. These are also visible in AC susceptibility [Fig. \ref{FIGURE1}(e)] without frequency dependence but with a thermal irreversibility at T$_{N2}$ (6.9/6.5~K once heating/cooling) along with an opening of Field-Cooled and Zero-Field-Cooled curves in DC magnetization below T$_{N2}$ \cite{SupplementalMaterial}, suggesting a first order transition. These results are consistent with previous findings \cite{Blasse1963,Strobel1980,Barton2014,Perversi2018,Zhou2023}, the small discrepancies related to the ordering temperatures or Curie-Weiss parameters being possibly due to antisite disorder allowing the presence of Fe$^{3+}$ \cite{SupplementalMaterial}. 

This double magnetic transition was investigated from single crystal ND on the CRG-D23 diffractometer at the Institut Laue-Langevin (ILL) \cite{SupplementalMaterial}. Figure \ref{FIGURE1}(f) shows the temperature evolution of the magnetic peaks at $Q$~=~(1~1~$\bar{1}$)-$q$ with the propagation vector $q$~=~($\delta$~$\delta$~0), characteristic of the magnetic order periodicity. A first incommensurate magnetic peak appears below T$_{N1}$~=~8.5~K with $\delta$~$\sim$~0.64~(r.l.u.), moves toward lower $\delta$ values once cooling and then disappears at 6~K. A second magnetic peak appears at 7.9~K (specific heat T$_{N2}$), first slightly incommensurate, before locking at the commensurate position $\delta$ = $\frac{2}{3}$ at 6.9~K (magnetization T$_{N2}$). The coexistence of both peaks confirms the first order nature of this second transition.

To proceed further in the identification of the magnetic structure, we used a multi-methods approach based on a combination of single crystal INS, ND and modeling \cite{SupplementalMaterial}. INS data presented in Fig. \ref{FIGURE2} were measured on the IN5 time-of-flight spectrometer at the ILL. The scattered intensity recorded at 12~K in the correlated paramagnetic regime rises steeply from $Q$ = (1 1 1)-$q$ and (1 1 3)-$q$ with $q$ = ($\frac{2}{3}$ $\frac{2}{3}$ 0) at zero energy and extends beyond 2.5~meV [Fig. \ref{FIGURE2}(a)]. Remarkably, this inelastic intensity exhibits a well-structured pattern in $Q$-space forming a network of connected triangles that get broader with increasing energy as observed from constant energy-cuts of the [00$l$]-[$hh$0] scattering plane [Fig. \ref{FIGURE2}(b)]. This triangle-based pattern is characteristic of the diffuse scattering of a triangular lattice with first neighbor AFM interactions and edge length corresponding to the Fe third neighbor distance in GFO. 

\begin{figure}
\includegraphics[width=1.0\columnwidth]{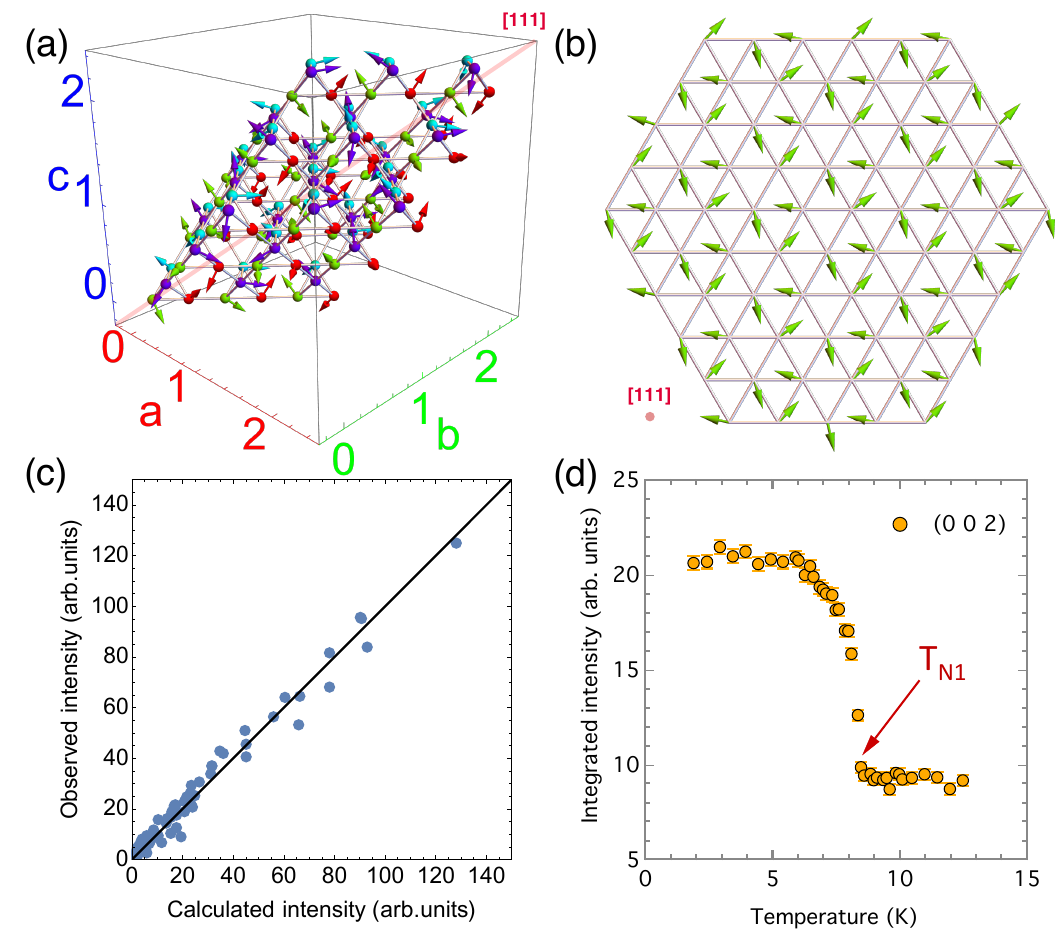}
\caption{(a) Sketch of the magnetic structure obtained with the model described in the text. The spins fill the rhombohedral primitive cell used in calculations, inserted in the cubic cell with the 4 Bravais lattices shown with different colors. (b) One triangular plane of a Bravais lattice viewed along the $[111]$ direction showing the 120$^{\circ}$ order. (c, d) ND results: agreement of the measured $vs$ calculated magnetic Bragg peak intensities (c) and temperature evolution of the integrated intensity of the forbidden (002) nuclear Bragg peak (d).}
\label{FIGURE3}
\end{figure}

Monte Carlo calculations were then performed above T$_{N1}$ by taking into account only the AFM $J_3$ interactions between easy-plane-like spins \cite{SupplementalMaterial}. They reproduce very well the triangular diffuse pattern [Fig. \ref{FIGURE2}(c)] demonstrating that the correlated regime corresponds to a 120$^{\circ}$ arrangement of the spins in the decoupled triangular planes built from dominant $J_3$. Adding other interactions modifies significantly this pattern \cite{SupplementalMaterial}. AFM $J_3'$ changes the bow-tie shapes of a triangle-pair into rectangles, while FM $J_1$ leads to some extinctions, $e.g.$ at $Q$ =($\pm$$\frac{2}{3}$, $\pm$$\frac{2}{3}$, 2). Note that these extinctions are observed in the ZnFe$_2$O$_4$ spinel whose FM $J_1$ interactions are therefore stronger and compete with $J_3$ \cite{Kamazawa2003}. In GFO, the deviation of the measured diffuse scattering from a triangular pattern is small, our calculations yielding the upper bound for the weaker $J_3'$ and $J_1$ interactions to 25\% of $\mid$$J_3$$\mid$ \cite{SupplementalMaterial}. Such observation indicates that there is a hierarchy of interactions in GFO, the third neighbor $J_3$ interactions being the dominant one \cite{Plumier1966}. This regime is therefore a good realization of a triangular antiferromagnet with some remarkable features: the first neighbor interactions of the triangular lattice are the pyrochlore third neighbors ones and the isolated triangular planes are actually distributed into 4 families of planes stacked along the cube's diagonals encompassing all the Fe ions. As the planes are decoupled above T$_{N1}$ while exhibiting strong magnetic correlations, we label this triangular correlated paramagnetic state a hierarchical spin liquid.

Having clearly identified the dominant interaction, we now focus on the mechanisms at the origin of the low temperature magnetic ordering. Using a dominant $J_3$ interaction as a starting point, we intend to find the Hamiltonian able to reproduce the observed ordered magnetic ground state by varying the other parameters. Using the Luttinger-Tisza-Bertaut method \cite{SupplementalMaterial}, $J_1$-$J_3'$ phase diagrams were calculated with fixed $J_3$ (AFM), an easy-plane anisotropy of $g_{\parallel}$/$g_{\perp}$ = 0.5 according to our CEF analysis, and different $J_n$/$\lvert J_3 \lvert$ ratios \cite{SupplementalMaterial}. Note that $J_3'$ interactions connect the spins within the planes of the same $\langle 111 \rangle$ triangular family while the connection between the 4 Bravais sublattices and thus the 3D ordering can be triggered by $J_1$. We identified phase space regions where the propagation vector $q$ is ($\frac{2}{3}$ $\frac{2}{3}$ 0) [Fig. \ref{FIGURE2}(f)]. We tested the corresponding Hamiltonians by confronting the spin wave dispersions calculated in the linear approximation of the Holstein-Primakoff formulation to INS spectra measured in various $Q$-directions at 1.5~K. A relative good agreement between simulations and experiments is obtained for the following ratios: $J_3'$/$\lvert J_3 \lvert$ = -0.5, $J_6$/$\lvert J_3 \lvert$ = -0.2, $J_1$/$\lvert J_3 \lvert$ = 0.2, $J_2$ = $J_4$ = 0 and an easy-plane anisotropy $g_{\parallel}$/$g_{\perp}$ = 0.5 [Figs. \ref{FIGURE2}(e)-(g)]. More details about other tested models, also considering $J_2$ and $J_4$, and additional INS data are provided in the Supplemental Material \cite{SupplementalMaterial}. To complement this analysis and formally consider the magneto-crystalline anisotropy determined from the CEF analysis, we have also performed RPA calculations shown in the Supplemental Material \cite{SupplementalMaterial}. The best RPA model yields spin waves and $J_n$/$\lvert J_3 \lvert$ ratios similar to those obtained from the linear spin wave calculations with the final parameters: $J_3$ = -0.23 $\pm$ 0.02~meV, $J_3'$ = -0.11 $\pm$ 0.01~meV, $J_6$ = -0.05 $\pm$ 0.01~meV and $J_1$ = 0.05 $\pm$ 0.02~meV. The agreement between both approaches is a good indication of the robustness of our model. Although not included in our best model, $J_2$ or $J_4$ cannot be completely ruled out with estimated upper absolute values equal to 0.02~meV. The long-distance $J_6$ interaction (180° super-super-exchange path) is necessary to obtain the correct ground state, which seems quite generic in germanates \cite{Diaz2006}. 

The spin arrangement in GFO was obtained by real-space mean field energy minimization of Monte-Carlo spin configurations calculated with the above Hamiltonian. The validity of our model is confirmed by the good agreement between the calculated magnetic structure factor and the integrated intensities of the magnetic Bragg peaks measured with the D23 diffractometer [Fig. \ref{FIGURE3}(c)]. The only adjustable parameter is the ordered magnetic moment: $M=3.0(1)$~$\mu_{\mathrm{B}}$ \cite{SupplementalMaterial}. The stabilized structure is an original non-coplanar multi-$q$ structure, described by a combination of the 6 symmetry-equivalent propagation vectors. From the amplitude of their Fourier components (Table \ref{Table_1}), it is seen that, for each Bravais lattice, three play a major role while the three others are one order of magnitude smaller. The main Fourier components associated to each propagation vector are distributed by pairs among the 4 Bravais lattices. This encodes the 120° spins within each triangular plane, the interplane dephasing {\it i.e.} the rotation of the 120$^{\circ}$ patterns from plane to plane along their $\langle 111 \rangle$ perpendicular directions, and the spin arrangement between the Bravais lattices. Although quite complex at first glance [Fig. \ref{FIGURE3}(a)], the resulting magnetic structure shows an underlying simple order, inherited from the hierarchical spin liquid state, where the spins in each of the 4 families remain nearly confined in the triangular planes and at 120$^\circ$ from each others as shown in Fig. \ref{FIGURE3}(b) \cite{SupplementalMaterial}.

This unusual ordering is enabled by the $J_1$, $J_3'$ and $J_6$ interactions. Their values obtained in our best spin wave model at 1.5~K are however stronger than those suggested by the diffuse scattering analysis at 12~K in the hierarchical spin liquid state \cite{SupplementalMaterial}. This suggests a temperature dependence of these exchange paths which cannot arise solely from usual thermal expansion. We have therefore performed a ND experiment, which has evidenced a forbidden (002) nuclear peak, already weakly present at room temperature and rising strongly below T$_{N1}$ [Fig. \ref{FIGURE3}(d)] \cite{SupplementalMaterial}. While additional studies are needed to confirm that a structural phase transition with a proper symmetry lowering occurs at T$_{N1}$, such observations suggest that magneto-structural effects are at play in GFO. Whereas the exact ordering mechanism at the origin of the two close critical temperatures remains unclear, a structural distortion at T$_{N1}$ could play a role in reinforcing $J_1$ and $J_3'$, favoring the 3D ordering.

\begin{table}
\caption{Fourier component amplitudes of the 6-$q$ magnetic structure for the 4 Bravais lattices (details on the vector components are found in the Supplemental Material \cite{SupplementalMaterial}).\label{Table_1}}
\begin{ruledtabular}
    \resizebox{\columnwidth}{!}{
    \begin{tabular}{lcccc}

     Propagation  & $\mid$S($q$)$\mid$ Fe 1  & $\mid$S($q$)$\mid$ Fe 2 & $\mid$S($q$)$\mid$ Fe 3 & $\mid$S($q$)$\mid$ Fe 4 \\  
     vectors $q$&  (green)  & (blue)  & (purple)  & (red)  \\    
 \hline  \\
     (0, -$\frac{2}{3}$, $\frac{2}{3}$) & {\bf 1.29}  & {\bf 1.52} & 0.11 & 0.10 \\  

     (-$\frac{2}{3}$, 0, $\frac{2}{3}$) & {\bf 2.25}  & 0.19 & {\bf 2.71} & 0.21 \\

     ($\frac{2}{3}$, $\frac{2}{3}$, 0) & 0.17  & {\bf 1.93} & {\bf 1.93} & 0.17 \\  

     (-$\frac{2}{3}$, $\frac{2}{3}$, 0) & {\bf 2.59}  & 0.22 & 0.22 & {\bf 2.59} \\    

     ($\frac{2}{3}$, 0, $\frac{2}{3}$) & 0.21  & {\bf 2.71} & 0.19 & {\bf 2.25} \\    
 
     (0, -$\frac{2}{3}$, -$\frac{2}{3}$) & 0.10  & 0.11 & {\bf 1.52} & {\bf 1.29} \\   

    \end{tabular}}
\end{ruledtabular}
\end{table}

The GFO magnetic structure is interesting in two respects. First, it has been recently highlighted that frustration and competing interactions in AFM insulators can potentially generate the non-coplanar spin textures that are the expected building blocks in advanced magnetism \cite{Gobel2021}. Our finding confirms that spinels are a suitable platform for their discovery. A 3-$q$ hedgehog-like magnetic order has for instance been predicted in spinels from a frustrated Heisenberg model on the pyrochlore lattice \cite{Zhitomirsky2022}. In Ni and Co members of the Ge spinel family, a 2-$q$ order made of hexagon vortices has been evidenced \cite{Beauvois2024}. The corresponding Hamiltonian includes single-ion anisotropy, dominant FM $J_1$ and AFM $J_3$ interactions and weaker $J_3'$, $J_4$, and $J_6$ in line with the present study. Finally, in the MnSc$_2$S$_4$ spinel, where magnetic ions occupy the diamond lattice, a 3-$q$ order of fractionalized skyrmions appears under magnetic field \cite{Gao2020,Rosales2022} due to competing interactions beyond first neighbors. Strikingly, while these and other recently discovered spin textures consist of whirling objects with potential topological properties, the fascinating 6-$q$ magnetic order arising in GFO materialises a new kind of non-coplanar spin texture built from the 3-dimensional coupling of 4 families of triangular planes with 120$^{\circ}$ spin ordering. Accordingly, a second interest comes from this triangular physics in itself since it has been the focus of many studies at the heart of frustrated magnetism \cite{Jolicoeur1989,Capriotti1999,Zheng2006,Zhitomirsky2013,Kim2019}. The non-collinear magnetic order promotes magnon-magnon interactions and magnon decays leading to roton-like features, broadened excitation linewidth or continuum, even in the classical case \cite{Oh2013}. GFO could be an interesting system to investigate the persistence of these features in a 3-dimensionally coupled triangular antiferromagnetic planes.

In conclusion, using an original multi-methods approach combining experiments and modeling, we propose that a complex multi-$q$ magnetic order is stabilized in GFO, which, remarkably, is produced by the leading third neighbor AFM interactions in the pyrochlore lattice. The dominance of these interactions leads to a hierarchical spin liquid in the paramagnetic state, made of isolated correlated planes materializing triangular antiferromagnets. At the ordering temperature, magneto-structural effects play a role in reinforcing the other interactions, then coupling the triangular planes together while retaining their 120$^{\circ}$ order. We anticipate interesting physical properties associated with this original realization of triangular physics, connected in three dimensions in a non-coplanar manner \cite{Zhou2023}.

\begin{acknowledgments} 
We are grateful to J. Debray and J. Balay for assistance with the sample preparation and Laue measurements. We also thank P. Lachkar for assistance with the specific heat measurements. We thank P. Fouilloux for his help with the neutron diffraction experiments. We thank E. Eyraud for access to the PMag platform, D. Jegouso for his help with the optical measurements and C. Paulsen for the use of his magnetometer. Neutron scattering experiments have been performed at Institut Laue-Langevin and Laboratoire Léon Brillouin-Orphée. We acknowledge financial support from the Institut de Physique CNRS (Grant Emergence@INP 2023) and ANR Grant No. ANR-15-CE30-0004. We thank M. Songvilay for fruitful discussions.
\end{acknowledgments}

\bibliographystyle{apsrev4-2}
\bibliography{GeFe2O4_MAIN_bib_vdec24_good}

%apsrev4-2.bst 2019-01-14 (MD) hand-edited version of apsrev4-1.bst
%Control: key (0)
%Control: author (72) initials jnrlst
%Control: editor formatted (1) identically to author
%Control: production of article title (-1) disabled
%Control: page (0) single
%Control: year (1) truncated
%Control: production of eprint (0) enabled
\begin{thebibliography}{47}%
\makeatletter
\providecommand \@ifxundefined [1]{%
 \@ifx{#1\undefined}
}%
\providecommand \@ifnum [1]{%
 \ifnum #1\expandafter \@firstoftwo
 \else \expandafter \@secondoftwo
 \fi
}%
\providecommand \@ifx [1]{%
 \ifx #1\expandafter \@firstoftwo
 \else \expandafter \@secondoftwo
 \fi
}%
\providecommand \natexlab [1]{#1}%
\providecommand \enquote  [1]{``#1''}%
\providecommand \bibnamefont  [1]{#1}%
\providecommand \bibfnamefont [1]{#1}%
\providecommand \citenamefont [1]{#1}%
\providecommand \href@noop [0]{\@secondoftwo}%
\providecommand \href [0]{\begingroup \@sanitize@url \@href}%
\providecommand \@href[1]{\@@startlink{#1}\@@href}%
\providecommand \@@href[1]{\endgroup#1\@@endlink}%
\providecommand \@sanitize@url [0]{\catcode `\\12\catcode `\$12\catcode `\&12\catcode `\#12\catcode `\^12\catcode `\_12\catcode `\%12\relax}%
\providecommand \@@startlink[1]{}%
\providecommand \@@endlink[0]{}%
\providecommand \url  [0]{\begingroup\@sanitize@url \@url }%
\providecommand \@url [1]{\endgroup\@href {#1}{\urlprefix }}%
\providecommand \urlprefix  [0]{URL }%
\providecommand \Eprint [0]{\href }%
\providecommand \doibase [0]{https://doi.org/}%
\providecommand \selectlanguage [0]{\@gobble}%
\providecommand \bibinfo  [0]{\@secondoftwo}%
\providecommand \bibfield  [0]{\@secondoftwo}%
\providecommand \translation [1]{[#1]}%
\providecommand \BibitemOpen [0]{}%
\providecommand \bibitemStop [0]{}%
\providecommand \bibitemNoStop [0]{.\EOS\space}%
\providecommand \EOS [0]{\spacefactor3000\relax}%
\providecommand \BibitemShut  [1]{\csname bibitem#1\endcsname}%
\let\auto@bib@innerbib\@empty
%</preamble>
\bibitem [{\citenamefont {Moessner}\ and\ \citenamefont {Ramirez}(2006)}]{Moessner2006}%
  \BibitemOpen
  \bibfield  {author} {\bibinfo {author} {\bibfnamefont {R.}~\bibnamefont {Moessner}}\ and\ \bibinfo {author} {\bibfnamefont {A.~P.}\ \bibnamefont {Ramirez}},\ }\href {https://doi.org/10.1063/1.2186278} {\bibfield  {journal} {\bibinfo  {journal} {Physics Today}\ }\textbf {\bibinfo {volume} {59}},\ \bibinfo {pages} {24–29} (\bibinfo {year} {2006})}\BibitemShut {NoStop}%
\bibitem [{\citenamefont {Normand}(2009)}]{Normand2009}%
  \BibitemOpen
  \bibfield  {author} {\bibinfo {author} {\bibfnamefont {B.}~\bibnamefont {Normand}},\ }\href {https://doi.org/10.1080/00107510902850361} {\bibfield  {journal} {\bibinfo  {journal} {Contemporary Physics}\ }\textbf {\bibinfo {volume} {50}},\ \bibinfo {pages} {533–552} (\bibinfo {year} {2009})}\BibitemShut {NoStop}%
\bibitem [{\citenamefont {Balents}(2010)}]{Balents2010}%
  \BibitemOpen
  \bibfield  {author} {\bibinfo {author} {\bibfnamefont {L.}~\bibnamefont {Balents}},\ }\href {https://doi.org/10.1038/nature08917} {\bibfield  {journal} {\bibinfo  {journal} {Nature}\ }\textbf {\bibinfo {volume} {464}},\ \bibinfo {pages} {199–208} (\bibinfo {year} {2010})}\BibitemShut {NoStop}%
\bibitem [{\citenamefont {Reimers}(1992)}]{Reimers1992}%
  \BibitemOpen
  \bibfield  {author} {\bibinfo {author} {\bibfnamefont {J.~N.}\ \bibnamefont {Reimers}},\ }\href {https://doi.org/10.1103/PhysRevB.45.7287} {\bibfield  {journal} {\bibinfo  {journal} {Phys. Rev. B}\ }\textbf {\bibinfo {volume} {45}},\ \bibinfo {pages} {7287–7294} (\bibinfo {year} {1992})}\BibitemShut {NoStop}%
\bibitem [{\citenamefont {Canals}\ and\ \citenamefont {Lacroix}(1998)}]{Canals1998}%
  \BibitemOpen
  \bibfield  {author} {\bibinfo {author} {\bibfnamefont {B.}~\bibnamefont {Canals}}\ and\ \bibinfo {author} {\bibfnamefont {C.}~\bibnamefont {Lacroix}},\ }\href {https://doi.org/10.1103/PhysRevLett.80.2933} {\bibfield  {journal} {\bibinfo  {journal} {Phys. Rev. Lett.}\ }\textbf {\bibinfo {volume} {80}},\ \bibinfo {pages} {2933–2936} (\bibinfo {year} {1998})}\BibitemShut {NoStop}%
\bibitem [{\citenamefont {Moessner}\ and\ \citenamefont {Chalker}(1998)}]{Moessner1998}%
  \BibitemOpen
  \bibfield  {author} {\bibinfo {author} {\bibfnamefont {R.}~\bibnamefont {Moessner}}\ and\ \bibinfo {author} {\bibfnamefont {J.~T.}\ \bibnamefont {Chalker}},\ }\href {https://doi.org/10.1103/PhysRevLett.80.2929} {\bibfield  {journal} {\bibinfo  {journal} {Phys. Rev. Lett.}\ }\textbf {\bibinfo {volume} {80}},\ \bibinfo {pages} {2929–2932} (\bibinfo {year} {1998})}\BibitemShut {NoStop}%
\bibitem [{\citenamefont {Lee}\ \emph {et~al.}(2002)\citenamefont {Lee}, \citenamefont {Broholm}, \citenamefont {Ratcliff}, \citenamefont {Gasparovic}, \citenamefont {Huang}, \citenamefont {Kim},\ and\ \citenamefont {Cheong}}]{Lee2002}%
  \BibitemOpen
  \bibfield  {author} {\bibinfo {author} {\bibfnamefont {S.~H.}\ \bibnamefont {Lee}}, \bibinfo {author} {\bibfnamefont {C.}~\bibnamefont {Broholm}}, \bibinfo {author} {\bibfnamefont {W.}~\bibnamefont {Ratcliff}}, \bibinfo {author} {\bibfnamefont {G.}~\bibnamefont {Gasparovic}}, \bibinfo {author} {\bibfnamefont {Q.}~\bibnamefont {Huang}}, \bibinfo {author} {\bibfnamefont {T.~H.}\ \bibnamefont {Kim}},\ and\ \bibinfo {author} {\bibfnamefont {S.~W.}\ \bibnamefont {Cheong}},\ }\href {https://doi.org/10.1038/nature00964} {\bibfield  {journal} {\bibinfo  {journal} {Nature}\ }\textbf {\bibinfo {volume} {418}},\ \bibinfo {pages} {856} (\bibinfo {year} {2002})}\BibitemShut {NoStop}%
\bibitem [{\citenamefont {Gardner}\ \emph {et~al.}(1999)\citenamefont {Gardner}, \citenamefont {Dunsiger}, \citenamefont {Gaulin}, \citenamefont {Gingras}, \citenamefont {Greedan}, \citenamefont {Kiefl}, \citenamefont {Lumsden}, \citenamefont {MacFarlane}, \citenamefont {Raju}, \citenamefont {Sonier}, \citenamefont {Swainson},\ and\ \citenamefont {Tun}}]{Gardner1999}%
  \BibitemOpen
  \bibfield  {author} {\bibinfo {author} {\bibfnamefont {J.~S.}\ \bibnamefont {Gardner}}, \bibinfo {author} {\bibfnamefont {S.~R.}\ \bibnamefont {Dunsiger}}, \bibinfo {author} {\bibfnamefont {B.~D.}\ \bibnamefont {Gaulin}}, \bibinfo {author} {\bibfnamefont {M.~J.~P.}\ \bibnamefont {Gingras}}, \bibinfo {author} {\bibfnamefont {J.~E.}\ \bibnamefont {Greedan}}, \bibinfo {author} {\bibfnamefont {R.~F.}\ \bibnamefont {Kiefl}}, \bibinfo {author} {\bibfnamefont {M.~D.}\ \bibnamefont {Lumsden}}, \bibinfo {author} {\bibfnamefont {W.~A.}\ \bibnamefont {MacFarlane}}, \bibinfo {author} {\bibfnamefont {N.~P.}\ \bibnamefont {Raju}}, \bibinfo {author} {\bibfnamefont {J.~E.}\ \bibnamefont {Sonier}}, \bibinfo {author} {\bibfnamefont {I.}~\bibnamefont {Swainson}},\ and\ \bibinfo {author} {\bibfnamefont {Z.}~\bibnamefont {Tun}},\ }\href {https://doi.org/10.1103/PhysRevLett.82.1012} {\bibfield  {journal} {\bibinfo  {journal} {Phys. Rev. Lett.}\ }\textbf {\bibinfo {volume} {82}},\ \bibinfo {pages} {1012} (\bibinfo {year}
  {1999})}\BibitemShut {NoStop}%
\bibitem [{\citenamefont {Gingras}\ and\ \citenamefont {McClarty}(2014)}]{Gingras_McClarty_2014}%
  \BibitemOpen
  \bibfield  {author} {\bibinfo {author} {\bibfnamefont {M.~J.~P.}\ \bibnamefont {Gingras}}\ and\ \bibinfo {author} {\bibfnamefont {P.~A.}\ \bibnamefont {McClarty}},\ }\href {https://doi.org/10.1088/0034-4885/77/5/056501} {\bibfield  {journal} {\bibinfo  {journal} {Rep. Prog. Phys.}\ }\textbf {\bibinfo {volume} {77}},\ \bibinfo {pages} {056501} (\bibinfo {year} {2014})}\BibitemShut {NoStop}%
\bibitem [{\citenamefont {Udagawa}\ and\ \citenamefont {Jaubert}(2021)}]{Udagawa2021}%
  \BibitemOpen
  \bibinfo {editor} {\bibfnamefont {M.}~\bibnamefont {Udagawa}}\ and\ \bibinfo {editor} {\bibfnamefont {L.}~\bibnamefont {Jaubert}},\ eds.,\ \href {https://doi.org/10.1007/978-3-030-70860-3} {\emph {\bibinfo {title} {Spin Ice}}},\ \bibinfo {series} {Springer Series in Solid-State Sciences}, Vol.\ \bibinfo {volume} {197}\ (\bibinfo  {publisher} {Springer International Publishing},\ \bibinfo {address} {Cham},\ \bibinfo {year} {2021})\BibitemShut {NoStop}%
\bibitem [{\citenamefont {Gardner}\ \emph {et~al.}(2010)\citenamefont {Gardner}, \citenamefont {Gingras},\ and\ \citenamefont {Greedan}}]{Gardner2010}%
  \BibitemOpen
  \bibfield  {author} {\bibinfo {author} {\bibfnamefont {J.~S.}\ \bibnamefont {Gardner}}, \bibinfo {author} {\bibfnamefont {M.~J.~P.}\ \bibnamefont {Gingras}},\ and\ \bibinfo {author} {\bibfnamefont {J.~E.}\ \bibnamefont {Greedan}},\ }\href {https://doi.org/10.1103/RevModPhys.82.53} {\bibfield  {journal} {\bibinfo  {journal} {Rev. Mod. Phys.}\ }\textbf {\bibinfo {volume} {82}},\ \bibinfo {pages} {53–107} (\bibinfo {year} {2010})}\BibitemShut {NoStop}%
\bibitem [{\citenamefont {Tsurkan}\ \emph {et~al.}(2021)\citenamefont {Tsurkan}, \citenamefont {Krug~von Nidda}, \citenamefont {Deisenhofer}, \citenamefont {Lunkenheimer},\ and\ \citenamefont {Loidl}}]{Tsurkan2021}%
  \BibitemOpen
  \bibfield  {author} {\bibinfo {author} {\bibfnamefont {V.}~\bibnamefont {Tsurkan}}, \bibinfo {author} {\bibfnamefont {H.-A.}\ \bibnamefont {Krug~von Nidda}}, \bibinfo {author} {\bibfnamefont {J.}~\bibnamefont {Deisenhofer}}, \bibinfo {author} {\bibfnamefont {P.}~\bibnamefont {Lunkenheimer}},\ and\ \bibinfo {author} {\bibfnamefont {A.}~\bibnamefont {Loidl}},\ }\href {https://doi.org/10.1016/j.physrep.2021.04.002} {\bibfield  {journal} {\bibinfo  {journal} {Physics Reports}\ }\textbf {\bibinfo {volume} {926}},\ \bibinfo {pages} {1} (\bibinfo {year} {2021})}\BibitemShut {NoStop}%
\bibitem [{\citenamefont {Lee}\ \emph {et~al.}(2010)\citenamefont {Lee}, \citenamefont {Takagi}, \citenamefont {Louca}, \citenamefont {Matsuda}, \citenamefont {Ji}, \citenamefont {Ueda}, \citenamefont {Ueda}, \citenamefont {Katsufuji}, \citenamefont {Chung}, \citenamefont {Park}, \citenamefont {Cheong},\ and\ \citenamefont {Broholm}}]{Lee2010}%
  \BibitemOpen
  \bibfield  {author} {\bibinfo {author} {\bibfnamefont {S.-H.}\ \bibnamefont {Lee}}, \bibinfo {author} {\bibfnamefont {H.}~\bibnamefont {Takagi}}, \bibinfo {author} {\bibfnamefont {D.}~\bibnamefont {Louca}}, \bibinfo {author} {\bibfnamefont {M.}~\bibnamefont {Matsuda}}, \bibinfo {author} {\bibfnamefont {S.}~\bibnamefont {Ji}}, \bibinfo {author} {\bibfnamefont {H.}~\bibnamefont {Ueda}}, \bibinfo {author} {\bibfnamefont {Y.}~\bibnamefont {Ueda}}, \bibinfo {author} {\bibfnamefont {T.}~\bibnamefont {Katsufuji}}, \bibinfo {author} {\bibfnamefont {J.-H.}\ \bibnamefont {Chung}}, \bibinfo {author} {\bibfnamefont {S.}~\bibnamefont {Park}}, \bibinfo {author} {\bibfnamefont {S.-W.}\ \bibnamefont {Cheong}},\ and\ \bibinfo {author} {\bibfnamefont {C.}~\bibnamefont {Broholm}},\ }\href {https://doi.org/10.1143/JPSJ.79.011004} {\bibfield  {journal} {\bibinfo  {journal} {J. Phys. Soc. Jpn.}\ }\textbf {\bibinfo {volume} {79}},\ \bibinfo {pages} {011004} (\bibinfo {year} {2010})}\BibitemShut {NoStop}%
\bibitem [{\citenamefont {Ueda}\ \emph {et~al.}(2006)\citenamefont {Ueda}, \citenamefont {Mitamura}, \citenamefont {Goto},\ and\ \citenamefont {Ueda}}]{Ueda2006}%
  \BibitemOpen
  \bibfield  {author} {\bibinfo {author} {\bibfnamefont {H.}~\bibnamefont {Ueda}}, \bibinfo {author} {\bibfnamefont {H.}~\bibnamefont {Mitamura}}, \bibinfo {author} {\bibfnamefont {T.}~\bibnamefont {Goto}},\ and\ \bibinfo {author} {\bibfnamefont {Y.}~\bibnamefont {Ueda}},\ }\href {https://doi.org/10.1103/PhysRevB.73.094415} {\bibfield  {journal} {\bibinfo  {journal} {Phys. Rev. B}\ }\textbf {\bibinfo {volume} {73}},\ \bibinfo {pages} {094415} (\bibinfo {year} {2006})}\BibitemShut {NoStop}%
\bibitem [{\citenamefont {Chung}\ \emph {et~al.}(2005)\citenamefont {Chung}, \citenamefont {Matsuda}, \citenamefont {Lee}, \citenamefont {Kakurai}, \citenamefont {Ueda}, \citenamefont {Sato}, \citenamefont {Takagi}, \citenamefont {Hong},\ and\ \citenamefont {Park}}]{Chung2005}%
  \BibitemOpen
  \bibfield  {author} {\bibinfo {author} {\bibfnamefont {J.-H.}\ \bibnamefont {Chung}}, \bibinfo {author} {\bibfnamefont {M.}~\bibnamefont {Matsuda}}, \bibinfo {author} {\bibfnamefont {S.-H.}\ \bibnamefont {Lee}}, \bibinfo {author} {\bibfnamefont {K.}~\bibnamefont {Kakurai}}, \bibinfo {author} {\bibfnamefont {H.}~\bibnamefont {Ueda}}, \bibinfo {author} {\bibfnamefont {T.~J.}\ \bibnamefont {Sato}}, \bibinfo {author} {\bibfnamefont {H.}~\bibnamefont {Takagi}}, \bibinfo {author} {\bibfnamefont {K.-P.}\ \bibnamefont {Hong}},\ and\ \bibinfo {author} {\bibfnamefont {S.}~\bibnamefont {Park}},\ }\href {https://doi.org/10.1103/PhysRevLett.95.247204} {\bibfield  {journal} {\bibinfo  {journal} {Phys. Rev. Lett.}\ }\textbf {\bibinfo {volume} {95}},\ \bibinfo {pages} {247204} (\bibinfo {year} {2005})}\BibitemShut {NoStop}%
\bibitem [{\citenamefont {Chern}\ \emph {et~al.}(2006)\citenamefont {Chern}, \citenamefont {Fennie},\ and\ \citenamefont {Tchernyshyov}}]{Chern2006}%
  \BibitemOpen
  \bibfield  {author} {\bibinfo {author} {\bibfnamefont {G.-W.}\ \bibnamefont {Chern}}, \bibinfo {author} {\bibfnamefont {C.~J.}\ \bibnamefont {Fennie}},\ and\ \bibinfo {author} {\bibfnamefont {O.}~\bibnamefont {Tchernyshyov}},\ }\href {https://doi.org/10.1103/PhysRevB.74.060405} {\bibfield  {journal} {\bibinfo  {journal} {Phys. Rev. B}\ }\textbf {\bibinfo {volume} {74}},\ \bibinfo {pages} {060405} (\bibinfo {year} {2006})}\BibitemShut {NoStop}%
\bibitem [{\citenamefont {Gao}\ \emph {et~al.}(2018)\citenamefont {Gao}, \citenamefont {Guratinder}, \citenamefont {Stuhr}, \citenamefont {White}, \citenamefont {Mansson}, \citenamefont {Roessli}, \citenamefont {Fennell}, \citenamefont {Tsurkan}, \citenamefont {Loidl}, \citenamefont {Ciomaga~Hatnean}, \citenamefont {Balakrishnan}, \citenamefont {Raymond}, \citenamefont {Chapon}, \citenamefont {Garlea}, \citenamefont {Savici}, \citenamefont {Cervellino}, \citenamefont {Bombardi}, \citenamefont {Chernyshov}, \citenamefont {R\"uegg}, \citenamefont {Haraldsen},\ and\ \citenamefont {Zaharko}}]{Gao2018}%
  \BibitemOpen
  \bibfield  {author} {\bibinfo {author} {\bibfnamefont {S.}~\bibnamefont {Gao}}, \bibinfo {author} {\bibfnamefont {K.}~\bibnamefont {Guratinder}}, \bibinfo {author} {\bibfnamefont {U.}~\bibnamefont {Stuhr}}, \bibinfo {author} {\bibfnamefont {J.~S.}\ \bibnamefont {White}}, \bibinfo {author} {\bibfnamefont {M.}~\bibnamefont {Mansson}}, \bibinfo {author} {\bibfnamefont {B.}~\bibnamefont {Roessli}}, \bibinfo {author} {\bibfnamefont {T.}~\bibnamefont {Fennell}}, \bibinfo {author} {\bibfnamefont {V.}~\bibnamefont {Tsurkan}}, \bibinfo {author} {\bibfnamefont {A.}~\bibnamefont {Loidl}}, \bibinfo {author} {\bibfnamefont {M.}~\bibnamefont {Ciomaga~Hatnean}}, \bibinfo {author} {\bibfnamefont {G.}~\bibnamefont {Balakrishnan}}, \bibinfo {author} {\bibfnamefont {S.}~\bibnamefont {Raymond}}, \bibinfo {author} {\bibfnamefont {L.}~\bibnamefont {Chapon}}, \bibinfo {author} {\bibfnamefont {V.~O.}\ \bibnamefont {Garlea}}, \bibinfo {author} {\bibfnamefont {A.~T.}\ \bibnamefont {Savici}}, \bibinfo {author} {\bibfnamefont
  {A.}~\bibnamefont {Cervellino}}, \bibinfo {author} {\bibfnamefont {A.}~\bibnamefont {Bombardi}}, \bibinfo {author} {\bibfnamefont {D.}~\bibnamefont {Chernyshov}}, \bibinfo {author} {\bibfnamefont {C.}~\bibnamefont {R\"uegg}}, \bibinfo {author} {\bibfnamefont {J.~T.}\ \bibnamefont {Haraldsen}},\ and\ \bibinfo {author} {\bibfnamefont {O.}~\bibnamefont {Zaharko}},\ }\href {https://doi.org/10.1103/PhysRevB.97.134430} {\bibfield  {journal} {\bibinfo  {journal} {Phys. Rev. B}\ }\textbf {\bibinfo {volume} {97}},\ \bibinfo {pages} {134430} (\bibinfo {year} {2018})}\BibitemShut {NoStop}%
\bibitem [{\citenamefont {Yaresko}(2008)}]{Yaresko2008}%
  \BibitemOpen
  \bibfield  {author} {\bibinfo {author} {\bibfnamefont {A.~N.}\ \bibnamefont {Yaresko}},\ }\href {https://doi.org/10.1103/PhysRevB.77.115106} {\bibfield  {journal} {\bibinfo  {journal} {Phys. Rev. B}\ }\textbf {\bibinfo {volume} {77}},\ \bibinfo {pages} {115106} (\bibinfo {year} {2008})}\BibitemShut {NoStop}%
\bibitem [{\citenamefont {Kamazawa}\ \emph {et~al.}(1999)\citenamefont {Kamazawa}, \citenamefont {Tsunoda}, \citenamefont {Odaka},\ and\ \citenamefont {Kohn}}]{Kamazawa1999}%
  \BibitemOpen
  \bibfield  {author} {\bibinfo {author} {\bibfnamefont {K.}~\bibnamefont {Kamazawa}}, \bibinfo {author} {\bibfnamefont {Y.}~\bibnamefont {Tsunoda}}, \bibinfo {author} {\bibfnamefont {K.}~\bibnamefont {Odaka}},\ and\ \bibinfo {author} {\bibfnamefont {K.}~\bibnamefont {Kohn}},\ }\href {https://doi.org/https://doi.org/10.1016/S0022-3697(99)00099-2} {\bibfield  {journal} {\bibinfo  {journal} {Journal of Physics and Chemistry of Solids}\ }\textbf {\bibinfo {volume} {60}},\ \bibinfo {pages} {1261} (\bibinfo {year} {1999})}\BibitemShut {NoStop}%
\bibitem [{\citenamefont {Kamazawa}\ \emph {et~al.}(2003)\citenamefont {Kamazawa}, \citenamefont {Tsunoda}, \citenamefont {Kadowaki},\ and\ \citenamefont {Kohn}}]{Kamazawa2003}%
  \BibitemOpen
  \bibfield  {author} {\bibinfo {author} {\bibfnamefont {K.}~\bibnamefont {Kamazawa}}, \bibinfo {author} {\bibfnamefont {Y.}~\bibnamefont {Tsunoda}}, \bibinfo {author} {\bibfnamefont {H.}~\bibnamefont {Kadowaki}},\ and\ \bibinfo {author} {\bibfnamefont {K.}~\bibnamefont {Kohn}},\ }\href {https://doi.org/10.1103/PhysRevB.68.024412} {\bibfield  {journal} {\bibinfo  {journal} {Phys. Rev. B}\ }\textbf {\bibinfo {volume} {68}},\ \bibinfo {pages} {024412} (\bibinfo {year} {2003})}\BibitemShut {NoStop}%
\bibitem [{\citenamefont {Kamazawa}\ \emph {et~al.}(2004)\citenamefont {Kamazawa}, \citenamefont {Park}, \citenamefont {Lee}, \citenamefont {Sato},\ and\ \citenamefont {Tsunoda}}]{Kamazawa2004}%
  \BibitemOpen
  \bibfield  {author} {\bibinfo {author} {\bibfnamefont {K.}~\bibnamefont {Kamazawa}}, \bibinfo {author} {\bibfnamefont {S.}~\bibnamefont {Park}}, \bibinfo {author} {\bibfnamefont {S.-H.}\ \bibnamefont {Lee}}, \bibinfo {author} {\bibfnamefont {T.~J.}\ \bibnamefont {Sato}},\ and\ \bibinfo {author} {\bibfnamefont {Y.}~\bibnamefont {Tsunoda}},\ }\href {https://doi.org/10.1103/PhysRevB.70.024418} {\bibfield  {journal} {\bibinfo  {journal} {Phys. Rev. B}\ }\textbf {\bibinfo {volume} {70}},\ \bibinfo {pages} {024418} (\bibinfo {year} {2004})}\BibitemShut {NoStop}%
\bibitem [{\citenamefont {Dronova}\ \emph {et~al.}(2024)\citenamefont {Dronova}, \citenamefont {Pet\ifmmode \check{r}\else \v{r}\fi{}\'{\i}\ifmmode~\check{c}\else \v{c}\fi{}ek}, \citenamefont {Morgan}, \citenamefont {Ye}, \citenamefont {Silevitch},\ and\ \citenamefont {Feng}}]{Dronova2024}%
  \BibitemOpen
  \bibfield  {author} {\bibinfo {author} {\bibfnamefont {M.~G.}\ \bibnamefont {Dronova}}, \bibinfo {author} {\bibfnamefont {V.}~\bibnamefont {Pet\ifmmode \check{r}\else \v{r}\fi{}\'{\i}\ifmmode~\check{c}\else \v{c}\fi{}ek}}, \bibinfo {author} {\bibfnamefont {Z.}~\bibnamefont {Morgan}}, \bibinfo {author} {\bibfnamefont {F.}~\bibnamefont {Ye}}, \bibinfo {author} {\bibfnamefont {D.~M.}\ \bibnamefont {Silevitch}},\ and\ \bibinfo {author} {\bibfnamefont {Y.}~\bibnamefont {Feng}},\ }\href {https://doi.org/10.1103/PhysRevB.109.064421} {\bibfield  {journal} {\bibinfo  {journal} {Phys. Rev. B}\ }\textbf {\bibinfo {volume} {109}},\ \bibinfo {pages} {064421} (\bibinfo {year} {2024})}\BibitemShut {NoStop}%
\bibitem [{\citenamefont {Goodenough}(1963)}]{Goodenough}%
  \BibitemOpen
  \bibfield  {author} {\bibinfo {author} {\bibfnamefont {J.~B.}\ \bibnamefont {Goodenough}},\ }\href@noop {} {\emph {\bibinfo {title} {Magnetism And The Chemical Bond}}}\ (\bibinfo  {publisher} {John Wiley And Sons},\ \bibinfo {year} {1963})\BibitemShut {NoStop}%
\bibitem [{\citenamefont {Perversi}\ \emph {et~al.}(2018)\citenamefont {Perversi}, \citenamefont {Arevalo-Lopez}, \citenamefont {Ritter},\ and\ \citenamefont {Attfield}}]{Perversi2018}%
  \BibitemOpen
  \bibfield  {author} {\bibinfo {author} {\bibfnamefont {G.}~\bibnamefont {Perversi}}, \bibinfo {author} {\bibfnamefont {A.~M.}\ \bibnamefont {Arevalo-Lopez}}, \bibinfo {author} {\bibfnamefont {C.}~\bibnamefont {Ritter}},\ and\ \bibinfo {author} {\bibfnamefont {J.~P.}\ \bibnamefont {Attfield}},\ }\href {https://doi.org/10.1038/s42005-018-0067-7} {\bibfield  {journal} {\bibinfo  {journal} {Communications Physics}\ }\textbf {\bibinfo {volume} {1}},\ \bibinfo {pages} {69} (\bibinfo {year} {2018})}\BibitemShut {NoStop}%
\bibitem [{\citenamefont {Barton}\ \emph {et~al.}(2014)\citenamefont {Barton}, \citenamefont {Kemei}, \citenamefont {Gaultois}, \citenamefont {Moffitt}, \citenamefont {Darago}, \citenamefont {Seshadri}, \citenamefont {Suchomel},\ and\ \citenamefont {Melot}}]{Barton2014}%
  \BibitemOpen
  \bibfield  {author} {\bibinfo {author} {\bibfnamefont {P.~T.}\ \bibnamefont {Barton}}, \bibinfo {author} {\bibfnamefont {M.~C.}\ \bibnamefont {Kemei}}, \bibinfo {author} {\bibfnamefont {M.~W.}\ \bibnamefont {Gaultois}}, \bibinfo {author} {\bibfnamefont {S.~L.}\ \bibnamefont {Moffitt}}, \bibinfo {author} {\bibfnamefont {L.~E.}\ \bibnamefont {Darago}}, \bibinfo {author} {\bibfnamefont {R.}~\bibnamefont {Seshadri}}, \bibinfo {author} {\bibfnamefont {M.~R.}\ \bibnamefont {Suchomel}},\ and\ \bibinfo {author} {\bibfnamefont {B.~C.}\ \bibnamefont {Melot}},\ }\href {https://doi.org/10.1103/PhysRevB.90.064105} {\bibfield  {journal} {\bibinfo  {journal} {Phys. Rev. B}\ }\textbf {\bibinfo {volume} {90}},\ \bibinfo {pages} {064105} (\bibinfo {year} {2014})}\BibitemShut {NoStop}%
\bibitem [{\citenamefont {Imbert}(1966)}]{Imbert1966}%
  \BibitemOpen
  \bibfield  {author} {\bibinfo {author} {\bibfnamefont {P.}~\bibnamefont {Imbert}},\ }\href@noop {} {\bibfield  {journal} {\bibinfo  {journal} {C. R. Acad. Sc. Paris s\'erie B}\ }\textbf {\bibinfo {volume} {263}},\ \bibinfo {pages} {184} (\bibinfo {year} {1966})}\BibitemShut {NoStop}%
\bibitem [{\citenamefont {Hartmann‐Boutron}\ and\ \citenamefont {Imbert}(1968)}]{Hartmann1968}%
  \BibitemOpen
  \bibfield  {author} {\bibinfo {author} {\bibfnamefont {F.}~\bibnamefont {Hartmann‐Boutron}}\ and\ \bibinfo {author} {\bibfnamefont {P.}~\bibnamefont {Imbert}},\ }\href {https://doi.org/10.1063/1.2163618} {\bibfield  {journal} {\bibinfo  {journal} {J. Appl. Phys.}\ }\textbf {\bibinfo {volume} {39}},\ \bibinfo {pages} {775} (\bibinfo {year} {1968})}\BibitemShut {NoStop}%
\bibitem [{\citenamefont {Plumier}(1966)}]{Plumier1966}%
  \BibitemOpen
  \bibfield  {author} {\bibinfo {author} {\bibfnamefont {R.}~\bibnamefont {Plumier}},\ }\href@noop {} {\bibfield  {journal} {\bibinfo  {journal} {C. R. Acad. Sc. Paris s\'erie B}\ }\textbf {\bibinfo {volume} {263}},\ \bibinfo {pages} {173} (\bibinfo {year} {1966})}\BibitemShut {NoStop}%
\bibitem [{\citenamefont {Sandemann}\ \emph {et~al.}(2023)\citenamefont {Sandemann}, \citenamefont {Gr{\o}nbech}, \citenamefont {St{\o}ckler}, \citenamefont {Ye}, \citenamefont {Chakoumakos},\ and\ \citenamefont {Iversen}}]{Sandemann2023}%
  \BibitemOpen
  \bibfield  {author} {\bibinfo {author} {\bibfnamefont {J.~R.}\ \bibnamefont {Sandemann}}, \bibinfo {author} {\bibfnamefont {T.~B.~E.}\ \bibnamefont {Gr{\o}nbech}}, \bibinfo {author} {\bibfnamefont {K.~A.~H.}\ \bibnamefont {St{\o}ckler}}, \bibinfo {author} {\bibfnamefont {F.}~\bibnamefont {Ye}}, \bibinfo {author} {\bibfnamefont {B.~C.}\ \bibnamefont {Chakoumakos}},\ and\ \bibinfo {author} {\bibfnamefont {B.~B.}\ \bibnamefont {Iversen}},\ }\href {https://doi.org/https://doi.org/10.1002/adma.202207152} {\bibfield  {journal} {\bibinfo  {journal} {Advanced Materials}\ }\textbf {\bibinfo {volume} {35}},\ \bibinfo {pages} {2207152} (\bibinfo {year} {2023})}\BibitemShut {NoStop}%
\bibitem [{\citenamefont {Matsuda}\ \emph {et~al.}(2008)\citenamefont {Matsuda}, \citenamefont {Chung}, \citenamefont {Park}, \citenamefont {Sato}, \citenamefont {Matsuno}, \citenamefont {Aruga~Katori}, \citenamefont {Takagi}, \citenamefont {Kakurai}, \citenamefont {Kamazawa}, \citenamefont {Tsunoda}, \citenamefont {Kagomiya}, \citenamefont {Henley},\ and\ \citenamefont {Lee}}]{Matsuda2008}%
  \BibitemOpen
  \bibfield  {author} {\bibinfo {author} {\bibfnamefont {M.}~\bibnamefont {Matsuda}}, \bibinfo {author} {\bibfnamefont {J.-H.}\ \bibnamefont {Chung}}, \bibinfo {author} {\bibfnamefont {S.}~\bibnamefont {Park}}, \bibinfo {author} {\bibfnamefont {T.~J.}\ \bibnamefont {Sato}}, \bibinfo {author} {\bibfnamefont {K.}~\bibnamefont {Matsuno}}, \bibinfo {author} {\bibfnamefont {H.}~\bibnamefont {Aruga~Katori}}, \bibinfo {author} {\bibfnamefont {H.}~\bibnamefont {Takagi}}, \bibinfo {author} {\bibfnamefont {K.}~\bibnamefont {Kakurai}}, \bibinfo {author} {\bibfnamefont {K.}~\bibnamefont {Kamazawa}}, \bibinfo {author} {\bibfnamefont {Y.}~\bibnamefont {Tsunoda}}, \bibinfo {author} {\bibfnamefont {I.}~\bibnamefont {Kagomiya}}, \bibinfo {author} {\bibfnamefont {C.~L.}\ \bibnamefont {Henley}},\ and\ \bibinfo {author} {\bibfnamefont {S.-H.}\ \bibnamefont {Lee}},\ }\href {https://doi.org/10.1209/0295-5075/82/37006} {\bibfield  {journal} {\bibinfo  {journal} {Europhysics Letters}\ }\textbf {\bibinfo {volume} {82}},\ \bibinfo
  {pages} {37006} (\bibinfo {year} {2008})}\BibitemShut {NoStop}%
\bibitem [{\citenamefont {Tomiyasu}\ and\ \citenamefont {Kamazawa}(2011)}]{Tomiyasu2011}%
  \BibitemOpen
  \bibfield  {author} {\bibinfo {author} {\bibfnamefont {K.}~\bibnamefont {Tomiyasu}}\ and\ \bibinfo {author} {\bibfnamefont {K.}~\bibnamefont {Kamazawa}},\ }\href {https://doi.org/10.1143/JPSJS.80SB.SB024} {\bibfield  {journal} {\bibinfo  {journal} {J. Phys. Soc. Jpn.}\ }\textbf {\bibinfo {volume} {80}},\ \bibinfo {pages} {SB024} (\bibinfo {year} {2011})}\BibitemShut {NoStop}%
\bibitem [{Sup()}]{SupplementalMaterial}%
  \BibitemOpen
  \href@noop {} {\bibinfo {title} {See {S}upplemental {M}aterial for additional information about the methods and supplementary data, analysis and calculations.}}\BibitemShut {Stop}%
\bibitem [{\citenamefont {Blasse}\ and\ \citenamefont {Fast}(1963)}]{Blasse1963}%
  \BibitemOpen
  \bibfield  {author} {\bibinfo {author} {\bibfnamefont {G.}~\bibnamefont {Blasse}}\ and\ \bibinfo {author} {\bibfnamefont {J.}~\bibnamefont {Fast}},\ }\href@noop {} {\bibfield  {journal} {\bibinfo  {journal} {Philips Research Reports}\ }\textbf {\bibinfo {volume} {18}},\ \bibinfo {pages} {393} (\bibinfo {year} {1963})}\BibitemShut {NoStop}%
\bibitem [{\citenamefont {Strobel}\ \emph {et~al.}(1980)\citenamefont {Strobel}, \citenamefont {Koffyberg},\ and\ \citenamefont {Wold}}]{Strobel1980}%
  \BibitemOpen
  \bibfield  {author} {\bibinfo {author} {\bibfnamefont {P.}~\bibnamefont {Strobel}}, \bibinfo {author} {\bibfnamefont {F.~P.}\ \bibnamefont {Koffyberg}},\ and\ \bibinfo {author} {\bibfnamefont {A.}~\bibnamefont {Wold}},\ }\href {https://doi.org/https://doi.org/10.1016/0022-4596(80)90022-5} {\bibfield  {journal} {\bibinfo  {journal} {Journal of Solid State Chemistry}\ }\textbf {\bibinfo {volume} {31}},\ \bibinfo {pages} {209} (\bibinfo {year} {1980})}\BibitemShut {NoStop}%
\bibitem [{\citenamefont {Zhou}\ \emph {et~al.}(2023)\citenamefont {Zhou}, \citenamefont {Tang}, \citenamefont {Lin}, \citenamefont {Huang}, \citenamefont {Zhang}, \citenamefont {Tang}, \citenamefont {Chen}, \citenamefont {Liu}, \citenamefont {Xie}, \citenamefont {Chen}, \citenamefont {Zheng}, \citenamefont {Yan}, \citenamefont {Jiang},\ and\ \citenamefont {Liu}}]{Zhou2023}%
  \BibitemOpen
  \bibfield  {author} {\bibinfo {author} {\bibfnamefont {G.}~\bibnamefont {Zhou}}, \bibinfo {author} {\bibfnamefont {Y.}~\bibnamefont {Tang}}, \bibinfo {author} {\bibfnamefont {L.}~\bibnamefont {Lin}}, \bibinfo {author} {\bibfnamefont {L.}~\bibnamefont {Huang}}, \bibinfo {author} {\bibfnamefont {J.}~\bibnamefont {Zhang}}, \bibinfo {author} {\bibfnamefont {Y.}~\bibnamefont {Tang}}, \bibinfo {author} {\bibfnamefont {P.}~\bibnamefont {Chen}}, \bibinfo {author} {\bibfnamefont {M.}~\bibnamefont {Liu}}, \bibinfo {author} {\bibfnamefont {Y.}~\bibnamefont {Xie}}, \bibinfo {author} {\bibfnamefont {X.}~\bibnamefont {Chen}}, \bibinfo {author} {\bibfnamefont {S.}~\bibnamefont {Zheng}}, \bibinfo {author} {\bibfnamefont {Z.}~\bibnamefont {Yan}}, \bibinfo {author} {\bibfnamefont {X.}~\bibnamefont {Jiang}},\ and\ \bibinfo {author} {\bibfnamefont {J.-M.}\ \bibnamefont {Liu}},\ }\href {https://doi.org/10.1088/1367-2630/ad131c} {\bibfield  {journal} {\bibinfo  {journal} {New J. Phys.}\ }\textbf {\bibinfo {volume} {25}},\
  \bibinfo {pages} {123033} (\bibinfo {year} {2023})}\BibitemShut {NoStop}%
\bibitem [{\citenamefont {Diaz}\ \emph {et~al.}(2006)\citenamefont {Diaz}, \citenamefont {de~Brion}, \citenamefont {Chouteau}, \citenamefont {Canals}, \citenamefont {Simonet},\ and\ \citenamefont {Strobel}}]{Diaz2006}%
  \BibitemOpen
  \bibfield  {author} {\bibinfo {author} {\bibfnamefont {S.}~\bibnamefont {Diaz}}, \bibinfo {author} {\bibfnamefont {S.}~\bibnamefont {de~Brion}}, \bibinfo {author} {\bibfnamefont {G.}~\bibnamefont {Chouteau}}, \bibinfo {author} {\bibfnamefont {B.}~\bibnamefont {Canals}}, \bibinfo {author} {\bibfnamefont {V.}~\bibnamefont {Simonet}},\ and\ \bibinfo {author} {\bibfnamefont {P.}~\bibnamefont {Strobel}},\ }\href {https://doi.org/10.1103/PhysRevB.74.092404} {\bibfield  {journal} {\bibinfo  {journal} {Phys. Rev. B}\ }\textbf {\bibinfo {volume} {74}},\ \bibinfo {pages} {092404} (\bibinfo {year} {2006})}\BibitemShut {NoStop}%
\bibitem [{\citenamefont {G{\"o}bel}\ \emph {et~al.}(2021)\citenamefont {G{\"o}bel}, \citenamefont {Mertig},\ and\ \citenamefont {Tretiakov}}]{Gobel2021}%
  \BibitemOpen
  \bibfield  {author} {\bibinfo {author} {\bibfnamefont {B.}~\bibnamefont {G{\"o}bel}}, \bibinfo {author} {\bibfnamefont {I.}~\bibnamefont {Mertig}},\ and\ \bibinfo {author} {\bibfnamefont {O.~A.}\ \bibnamefont {Tretiakov}},\ }\href {https://doi.org/https://doi.org/10.1016/j.physrep.2020.10.001} {\bibfield  {journal} {\bibinfo  {journal} {Physics Reports}\ }\textbf {\bibinfo {volume} {895}},\ \bibinfo {pages} {1} (\bibinfo {year} {2021})}\BibitemShut {NoStop}%
\bibitem [{\citenamefont {Zhitomirsky}\ \emph {et~al.}(2022)\citenamefont {Zhitomirsky}, \citenamefont {Gvozdikova},\ and\ \citenamefont {Ziman}}]{Zhitomirsky2022}%
  \BibitemOpen
  \bibfield  {author} {\bibinfo {author} {\bibfnamefont {M.}~\bibnamefont {Zhitomirsky}}, \bibinfo {author} {\bibfnamefont {M.}~\bibnamefont {Gvozdikova}},\ and\ \bibinfo {author} {\bibfnamefont {T.}~\bibnamefont {Ziman}},\ }\href {https://doi.org/10.1016/j.aop.2022.169066} {\bibfield  {journal} {\bibinfo  {journal} {Annals of Physics}\ }\textbf {\bibinfo {volume} {447}},\ \bibinfo {pages} {169066} (\bibinfo {year} {2022})}\BibitemShut {NoStop}%
\bibitem [{\citenamefont {Beauvois}\ and\ \citenamefont {\textit{et al.}}(2025)}]{Beauvois2024}%
  \BibitemOpen
  \bibfield  {author} {\bibinfo {author} {\bibfnamefont {K.}~\bibnamefont {Beauvois}}\ and\ \bibinfo {author} {\bibnamefont {\textit{et al.}}}} (\bibinfo {year} {2025}),\ \bibinfo {note} {{I}n preparation}\BibitemShut {NoStop}%
\bibitem [{\citenamefont {Gao}\ \emph {et~al.}(2020)\citenamefont {Gao}, \citenamefont {Rosales}, \citenamefont {G{\'o}mez~Albarrac{\'i}n}, \citenamefont {Tsurkan}, \citenamefont {Kaur}, \citenamefont {Fennell}, \citenamefont {Steffens}, \citenamefont {Boehm}, \citenamefont {{\v{C}}erm{\'a}k}, \citenamefont {Schneidewind}, \citenamefont {Ressouche}, \citenamefont {Cabra}, \citenamefont {R{\"u}egg},\ and\ \citenamefont {Zaharko}}]{Gao2020}%
  \BibitemOpen
  \bibfield  {author} {\bibinfo {author} {\bibfnamefont {S.}~\bibnamefont {Gao}}, \bibinfo {author} {\bibfnamefont {H.~D.}\ \bibnamefont {Rosales}}, \bibinfo {author} {\bibfnamefont {F.~A.}\ \bibnamefont {G{\'o}mez~Albarrac{\'i}n}}, \bibinfo {author} {\bibfnamefont {V.}~\bibnamefont {Tsurkan}}, \bibinfo {author} {\bibfnamefont {G.}~\bibnamefont {Kaur}}, \bibinfo {author} {\bibfnamefont {T.}~\bibnamefont {Fennell}}, \bibinfo {author} {\bibfnamefont {P.}~\bibnamefont {Steffens}}, \bibinfo {author} {\bibfnamefont {M.}~\bibnamefont {Boehm}}, \bibinfo {author} {\bibfnamefont {P.}~\bibnamefont {{\v{C}}erm{\'a}k}}, \bibinfo {author} {\bibfnamefont {A.}~\bibnamefont {Schneidewind}}, \bibinfo {author} {\bibfnamefont {E.}~\bibnamefont {Ressouche}}, \bibinfo {author} {\bibfnamefont {D.~C.}\ \bibnamefont {Cabra}}, \bibinfo {author} {\bibfnamefont {C.}~\bibnamefont {R{\"u}egg}},\ and\ \bibinfo {author} {\bibfnamefont {O.}~\bibnamefont {Zaharko}},\ }\href {https://doi.org/10.1038/s41586-020-2716-8} {\bibfield  {journal}
  {\bibinfo  {journal} {Nature}\ }\textbf {\bibinfo {volume} {586}},\ \bibinfo {pages} {37} (\bibinfo {year} {2020})}\BibitemShut {NoStop}%
\bibitem [{\citenamefont {Rosales}\ \emph {et~al.}(2022)\citenamefont {Rosales}, \citenamefont {Albarrac\'{\i}n}, \citenamefont {Guratinder}, \citenamefont {Tsurkan}, \citenamefont {Prodan}, \citenamefont {Ressouche},\ and\ \citenamefont {Zaharko}}]{Rosales2022}%
  \BibitemOpen
  \bibfield  {author} {\bibinfo {author} {\bibfnamefont {H.~D.}\ \bibnamefont {Rosales}}, \bibinfo {author} {\bibfnamefont {F.~A.~G.}\ \bibnamefont {Albarrac\'{\i}n}}, \bibinfo {author} {\bibfnamefont {K.}~\bibnamefont {Guratinder}}, \bibinfo {author} {\bibfnamefont {V.}~\bibnamefont {Tsurkan}}, \bibinfo {author} {\bibfnamefont {L.}~\bibnamefont {Prodan}}, \bibinfo {author} {\bibfnamefont {E.}~\bibnamefont {Ressouche}},\ and\ \bibinfo {author} {\bibfnamefont {O.}~\bibnamefont {Zaharko}},\ }\href {https://doi.org/10.1103/PhysRevB.105.224402} {\bibfield  {journal} {\bibinfo  {journal} {Phys. Rev. B}\ }\textbf {\bibinfo {volume} {105}},\ \bibinfo {pages} {224402} (\bibinfo {year} {2022})}\BibitemShut {NoStop}%
\bibitem [{\citenamefont {Jolicoeur}\ and\ \citenamefont {Le~Guillou}(1989)}]{Jolicoeur1989}%
  \BibitemOpen
  \bibfield  {author} {\bibinfo {author} {\bibfnamefont {T.}~\bibnamefont {Jolicoeur}}\ and\ \bibinfo {author} {\bibfnamefont {J.~C.}\ \bibnamefont {Le~Guillou}},\ }\href {https://doi.org/10.1103/PhysRevB.40.2727} {\bibfield  {journal} {\bibinfo  {journal} {Phys. Rev. B}\ }\textbf {\bibinfo {volume} {40}},\ \bibinfo {pages} {2727} (\bibinfo {year} {1989})}\BibitemShut {NoStop}%
\bibitem [{\citenamefont {Capriotti}\ \emph {et~al.}(1999)\citenamefont {Capriotti}, \citenamefont {Trumper},\ and\ \citenamefont {Sorella}}]{Capriotti1999}%
  \BibitemOpen
  \bibfield  {author} {\bibinfo {author} {\bibfnamefont {L.}~\bibnamefont {Capriotti}}, \bibinfo {author} {\bibfnamefont {A.~E.}\ \bibnamefont {Trumper}},\ and\ \bibinfo {author} {\bibfnamefont {S.}~\bibnamefont {Sorella}},\ }\href {https://doi.org/10.1103/PhysRevLett.82.3899} {\bibfield  {journal} {\bibinfo  {journal} {Phys. Rev. Lett.}\ }\textbf {\bibinfo {volume} {82}},\ \bibinfo {pages} {3899} (\bibinfo {year} {1999})}\BibitemShut {NoStop}%
\bibitem [{\citenamefont {Zheng}\ \emph {et~al.}(2006)\citenamefont {Zheng}, \citenamefont {Fj\ae{}restad}, \citenamefont {Singh}, \citenamefont {McKenzie},\ and\ \citenamefont {Coldea}}]{Zheng2006}%
  \BibitemOpen
  \bibfield  {author} {\bibinfo {author} {\bibfnamefont {W.}~\bibnamefont {Zheng}}, \bibinfo {author} {\bibfnamefont {J.~O.}\ \bibnamefont {Fj\ae{}restad}}, \bibinfo {author} {\bibfnamefont {R.~R.~P.}\ \bibnamefont {Singh}}, \bibinfo {author} {\bibfnamefont {R.~H.}\ \bibnamefont {McKenzie}},\ and\ \bibinfo {author} {\bibfnamefont {R.}~\bibnamefont {Coldea}},\ }\href {https://doi.org/10.1103/PhysRevLett.96.057201} {\bibfield  {journal} {\bibinfo  {journal} {Phys. Rev. Lett.}\ }\textbf {\bibinfo {volume} {96}},\ \bibinfo {pages} {057201} (\bibinfo {year} {2006})}\BibitemShut {NoStop}%
\bibitem [{\citenamefont {Zhitomirsky}\ and\ \citenamefont {Chernyshev}(2013)}]{Zhitomirsky2013}%
  \BibitemOpen
  \bibfield  {author} {\bibinfo {author} {\bibfnamefont {M.~E.}\ \bibnamefont {Zhitomirsky}}\ and\ \bibinfo {author} {\bibfnamefont {A.~L.}\ \bibnamefont {Chernyshev}},\ }\href {https://doi.org/10.1103/RevModPhys.85.219} {\bibfield  {journal} {\bibinfo  {journal} {Rev. Mod. Phys.}\ }\textbf {\bibinfo {volume} {85}},\ \bibinfo {pages} {219} (\bibinfo {year} {2013})}\BibitemShut {NoStop}%
\bibitem [{\citenamefont {Kim}\ \emph {et~al.}(2019)\citenamefont {Kim}, \citenamefont {Park}, \citenamefont {Leiner},\ and\ \citenamefont {Park}}]{Kim2019}%
  \BibitemOpen
  \bibfield  {author} {\bibinfo {author} {\bibfnamefont {T.}~\bibnamefont {Kim}}, \bibinfo {author} {\bibfnamefont {K.}~\bibnamefont {Park}}, \bibinfo {author} {\bibfnamefont {J.~C.}\ \bibnamefont {Leiner}},\ and\ \bibinfo {author} {\bibfnamefont {J.-G.}\ \bibnamefont {Park}},\ }\href {https://doi.org/10.7566/JPSJ.88.081003} {\bibfield  {journal} {\bibinfo  {journal} {J. Phys. Soc. Jpn.}\ }\textbf {\bibinfo {volume} {88}},\ \bibinfo {pages} {081003} (\bibinfo {year} {2019})}\BibitemShut {NoStop}%
\bibitem [{\citenamefont {Oh}\ \emph {et~al.}(2013)\citenamefont {Oh}, \citenamefont {Le}, \citenamefont {Jeong}, \citenamefont {Lee}, \citenamefont {Woo}, \citenamefont {Song}, \citenamefont {Perring}, \citenamefont {Buyers}, \citenamefont {Cheong},\ and\ \citenamefont {Park}}]{Oh2013}%
  \BibitemOpen
  \bibfield  {author} {\bibinfo {author} {\bibfnamefont {J.}~\bibnamefont {Oh}}, \bibinfo {author} {\bibfnamefont {M.~D.}\ \bibnamefont {Le}}, \bibinfo {author} {\bibfnamefont {J.}~\bibnamefont {Jeong}}, \bibinfo {author} {\bibfnamefont {J.-h.}\ \bibnamefont {Lee}}, \bibinfo {author} {\bibfnamefont {H.}~\bibnamefont {Woo}}, \bibinfo {author} {\bibfnamefont {W.-Y.}\ \bibnamefont {Song}}, \bibinfo {author} {\bibfnamefont {T.~G.}\ \bibnamefont {Perring}}, \bibinfo {author} {\bibfnamefont {W.~J.~L.}\ \bibnamefont {Buyers}}, \bibinfo {author} {\bibfnamefont {S.-W.}\ \bibnamefont {Cheong}},\ and\ \bibinfo {author} {\bibfnamefont {J.-G.}\ \bibnamefont {Park}},\ }\href {https://doi.org/10.1103/PhysRevLett.111.257202} {\bibfield  {journal} {\bibinfo  {journal} {Phys. Rev. Lett.}\ }\textbf {\bibinfo {volume} {111}},\ \bibinfo {pages} {257202} (\bibinfo {year} {2013})}\BibitemShut {NoStop}%
\end{thebibliography}%

\end{document}